\documentclass[aps,twocolumn]{revtex4}
\usepackage{epsfig}
\usepackage{graphicx}
\begin{document}
\title{Scaling exponent of the maximum
  growth probability in diffusion-limited aggregation}
\author {Mogens H. Jensen$^*$, Joachim Mathiesen$^*$ and Itamar Procaccia$^\dag$}
\affiliation{$^*$The Niels Bohr Institute, Blegdamsvej 17, Copenhagen, Denmark.\\
$^\dag$ Dept. of Chemical Physics, The Weizmann Institute of Science, Rehovot 76100, Israel.}
\begin{abstract}
An early (and influential) scaling relation in the multifractal theory of Diffusion Limited Aggregation
(DLA) is the Turkevich-Scher conjecture that relates the exponent $\alpha_{min}$ that characterizes
the ``hottest'' region of the harmonic measure and the fractal dimension D of the cluster, i.e.
$D=1+\alpha_{min}$. Due to lack of accurate direct measurements of both $D$ and $\alpha_{min}$
this conjecture could never be put to serious test. 
Using the method of iterated conformal
maps $D$ was recently determined as $D=1.713\pm 0.003$. In this Letter
we determine $\alpha_{min}$ accurately, with the 
result $\alpha_{min}=0.665\pm 0.004$. We thus conclude
that the Turkevich-Scher conjecture is incorrect for DLA. 
\end{abstract}
\maketitle

Multifractal measures are normalized distributions lying upon fractal sets.
As such, they present rich scaling properties that have attracted considerable
attention in the last two decades. In this Letter we address the harmonic measure of
Diffusion Limited Aggregates \cite{81WS}, which is the probability measure for a random
walker coming from infinity to hit the boundary of the fractal cluster. This was one of
the earliest multifractal measures to be studied in the physics literature \cite{86HMP}, 
but the elucidation of its properties was made difficult by the extreme
variation of the probability to hit the tips of a DLA versus hitting the deep fjords. 
Thus the understanding of its scaling properties has been a long standing
issue. These scaling properties are conveniently
studied using the notion of generalized dimensions $D_q$, and the
associated $f(\alpha)$ function \cite{83HP,86HJKSP}. The simplest definition of the
generalized dimensions
is in terms of a uniform covering of the boundary of a DLA cluster
with boxes of size $\ell$, and measuring the probability for a random
walker coming from infinity to hit a piece of boundary which belongs to
the $i$'th box.
Denoting this probability by $P_i(\ell)$, one considers \cite{83HP}
\begin{equation}
D_q \equiv \lim_{\ell \to 0}\frac{1}{q-1}\frac{\log\sum_i
P_i^q(\ell)}{\log\ell} \ ,
\end{equation}
where the index $i$ runs over all the boxes that contain a piece of the boundary.
The limit $D_0\equiv \lim_{q\to 0^+} D_q$ is the fractal, or box dimension of the cluster.
$D_1\equiv \lim_{q\to 1^+} D_q$ and
$D_2$ are the well known information and correlation 
dimensions respectively \cite{56BR,82Far,83GP}. It is
well established by now \cite{86HJKSP} that the existence of an interesting spectrum of
values $D_q$ is related to the probabilities $P_i(\ell)$ having a spectrum of
``singularities" in the
sense that $P_i(\ell) \sim \ell^\alpha$
with $\alpha$ taking on values from a range $\alpha_{\rm min}\le
\alpha\le \alpha_{\rm max}$.
The frequency of observation of a particular value of $\alpha$ is
determined by the function
$f(\alpha)$ where (with $\tau(q)\equiv (q-1)D_q$)
\begin{equation}
f(\alpha) = \alpha q(\alpha)-\tau\Big(q\left(\alpha\right)\Big)\ ,\quad
\frac{\partial \tau(q)}{\partial q}=\alpha(q) \ .
\end{equation}

Of particular interest are the values of the minimal
and maximal values, $\alpha_{min}$ and $\alpha_{max}$, relating
to the largest and smallest growth probabilities, respectively. The maximal
value $\alpha_{max}$ was a subject of a long controversy that was
settled only recently (cf. \cite{01DJLMP,02JLMP} and references therein).
The issue of $\alpha_{min}$ appears to be one of the last of the multifractal
properties of DLA that has resisted settling. This is the subject of 
this Letter.

Consider DLA clusters containing $n$ particles of radius $\sqrt{\lambda_0}$, and
denote the radius of the minimal circle that contains the cluster as $R_n$. An
incoming random walker from infinity has some probability to hit
any of the existing particles of the cluster. Denote the maximal of these
probabilities as $p_{max}$. The average of this probabilities over many 
clusters of $n$ particles appears to scale as
\begin{equation}
  \label{pmax}
\langle p_{max}\rangle \sim  \left(\frac{\sqrt\lambda_0}{R_n}\right)^{\alpha_{min}}\sim
n^{-\alpha_{min}/D} \ , 
\end{equation}
where for the last step we have used the obvious scaling law $n\sim (R_n/\sqrt{\lambda_0})^{D}$.
Turkevich and Scher have made the plausible assumption that the position of the
cluster particle associated with $p_{max}$ is at the outermost tip of the cluster.
Thus, a scaling relation can be derived by stating that upon adding a new particle
to the cluster, $R_n$ will grow by one unit $\sqrt{\lambda_0}$ with probability $p_{max}$,
or will not grow at all with probability $1-p_{max}$. Then 
\begin{equation}
  \label{dRdn}
  \frac {dR}{dn}\sim \sqrt{\lambda_0}~ p_{max} \sim n^{1/D-1} \ ,
\end{equation}
where again the last step stems from the definition of the fractal dimension.
Equating the RHS of (\ref{pmax}) and (\ref{dRdn}) we get the Turkevich-Scher 
conjecture
\begin{equation}
D=1+\alpha_{min} \ . \label{TS}
\end{equation}
We will show here that this conjecture is incorrect simply because the 
position of maximal probability is {\em not} at the outermost tip
of the DLA cluster. In fact, one can introduce in analogy to Eq. (\ref{pmax})
a scaling law for the probability to hit the actual tip of the cluster 
(the particle which is furthest away from the origin), i.e.
\begin{equation}
  \label{ptip}
\langle p_{tip}\rangle \sim  \left(\frac{\sqrt\lambda_0}{R_n}\right)^{\alpha_{tip}}\sim
n^{-\alpha_{tip}/D} \ . 
\end{equation}
A scaling law 
\begin{equation}
D=1+\alpha_{tip} \ , \label{tip}
\end{equation} is then a tautology. We will show 
that for DLA $\alpha_{tip}>\alpha_{min}$.

To achieve accurate estimates of $\alpha_{min}$ (and in passing of $\alpha_{tip}$) we resort
to the method of iterated conformal maps that was shown to be extremely
useful for dealing with DLA and related growth processes. The method
was amply described before, so we just remind the reader that it
is based on compositions of fundamental
conformal maps
$\phi_{\lambda,\theta}$ which map the exterior of the unit circle 
to its exterior, except for a little bump at $e^{i\theta}$ of linear size
proportional to $\sqrt\lambda$. The composition of these mappings is
analogous to the aggregation of random walkers in the off-lattice DLA
model. We shall here use the mapping
introduced in \cite{98HL} which produces two square root singularities which
we refer to as the branch cuts, and the tip of the bump which we refer to as 
the micro tip. The dynamics is given by 
\begin{equation}
  \label{eq:1}
\Phi^{(n)}(w) = \Phi^{(n-1)}(\phi_{\lambda_{n},\theta_{n}}(w)) \ .
\end{equation}
where $\Phi^{(n)}$ maps the exterior of the unit circle to the exterior
of the cluster of $n$ bumps. 
The size of the $n$'th bump is controlled by the parameter $\lambda_n$
and in order to achieve particles of fixed size we have that, to leading 
order,
\begin{equation}
  \label{eq:2}
  \lambda_{n} = \frac{\lambda_0}{|{\Phi^{(n-1)}}' (e^{i \theta_n})|^2}. 
\end{equation}

Using the iterated conformal maps it is very easy to keep track of where the maximum
growth probability is located, and where the outermost tip is, as more particles are added. Let us
assume that at the ($n$-1)'th growth step the site with the largest
probability is located at the angle $\theta_{max}$ on the unit circle,
i.e. for all $\theta$ 
\begin{equation}
  \label{eq:7}
\frac 1 {|{\Phi^{(n-1)}}' (e^{i \theta_{max}})|}\geq \frac 1 {|{\Phi^{(n-1)}}' (e^{i \theta})|} 
\end{equation}
\begin{figure}
\centering
\includegraphics[width=.35\textwidth]{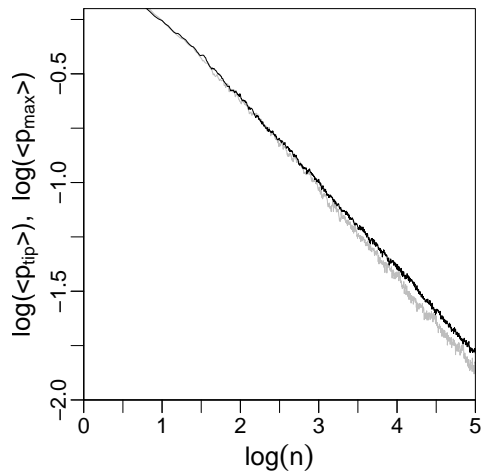}
\caption{The average of $p_{max}$ (upper,black line) and $p_{tip}$ (lower, grey line)
  versus $n$. The average is over $20$ clusters of sizes
  up to $n=100000$.  From the values of the slopes
  we estimate of $\alpha_{min}=0.681\pm.014$ and $\alpha_{tip}=0.713\pm.012$.}
\label{direct}
\end{figure}
When we add a new bump in the $n$'th growth step the position of maximal
probability may not change (up to reparameterization of the angle $\theta_{max}$),
or move to the new bump. We can easily find the reparameterized angle and determine
the new position from
\begin{equation}
  \label{pmaxn}
  p_{max,n}=\max\left\{\frac 1 {|{\Phi^{(n)}}'
(\phi^{-1}_{\lambda_n,\theta_n}(e^{i\theta_{max}}))|},\frac 1 {|{\Phi^{(n)}}' (e^{i
\theta_n})|}\right\}
\ .
\end{equation}
If $p_{max,n}$ is located at $\theta_n$ we put $\theta_{max}=\theta_n$
in the $(n+1)$'th growth step. Similarly we track the position $|z|_{max}$ on the
cluster by finding the value $\theta_{tip}$ which assign the maximal value of
$|\Phi^{(n)}(e^{i\theta})|$.  We compute
$p_{tip}$ there as $1/|{\Phi^{(n)}}'|$.

A direct measurement of $\alpha_{min}$ and $\alpha_{tip}$ is displayed
in Fig. \ref{direct}.  From the direct measurement we find $\alpha_{min}\approx 0.681$
while $\alpha_{tip}\approx 0.713$. Clearly the latter is in agreement with (\ref{tip})
while the former is in disagreement with (\ref{TS}) (taking as a datum
the result of \cite{00DLP}, $D=1.713\pm 0.005$).

\begin{figure}
\centering
\includegraphics[width=.35\textwidth]{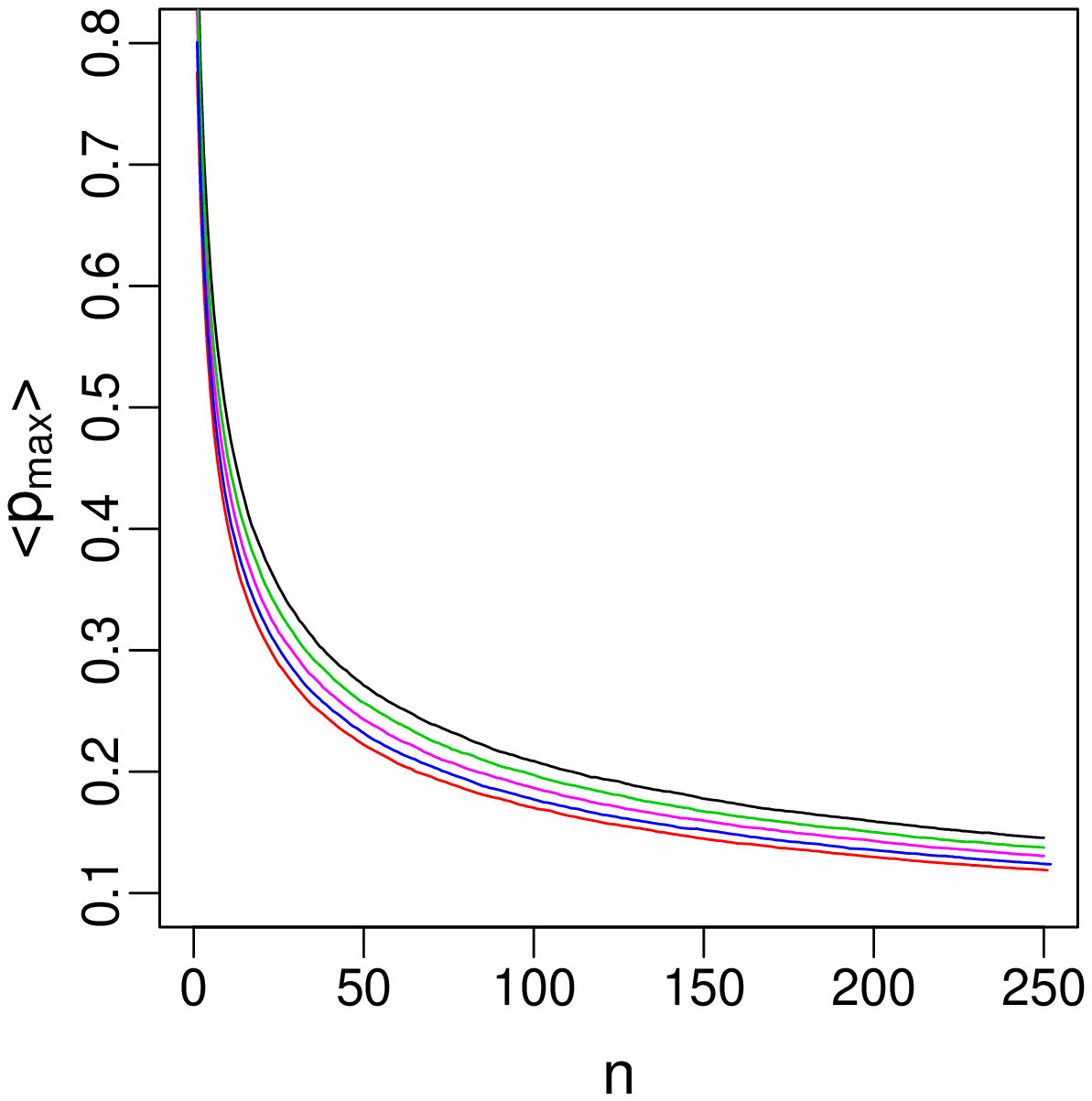}
\includegraphics[width=.35\textwidth]{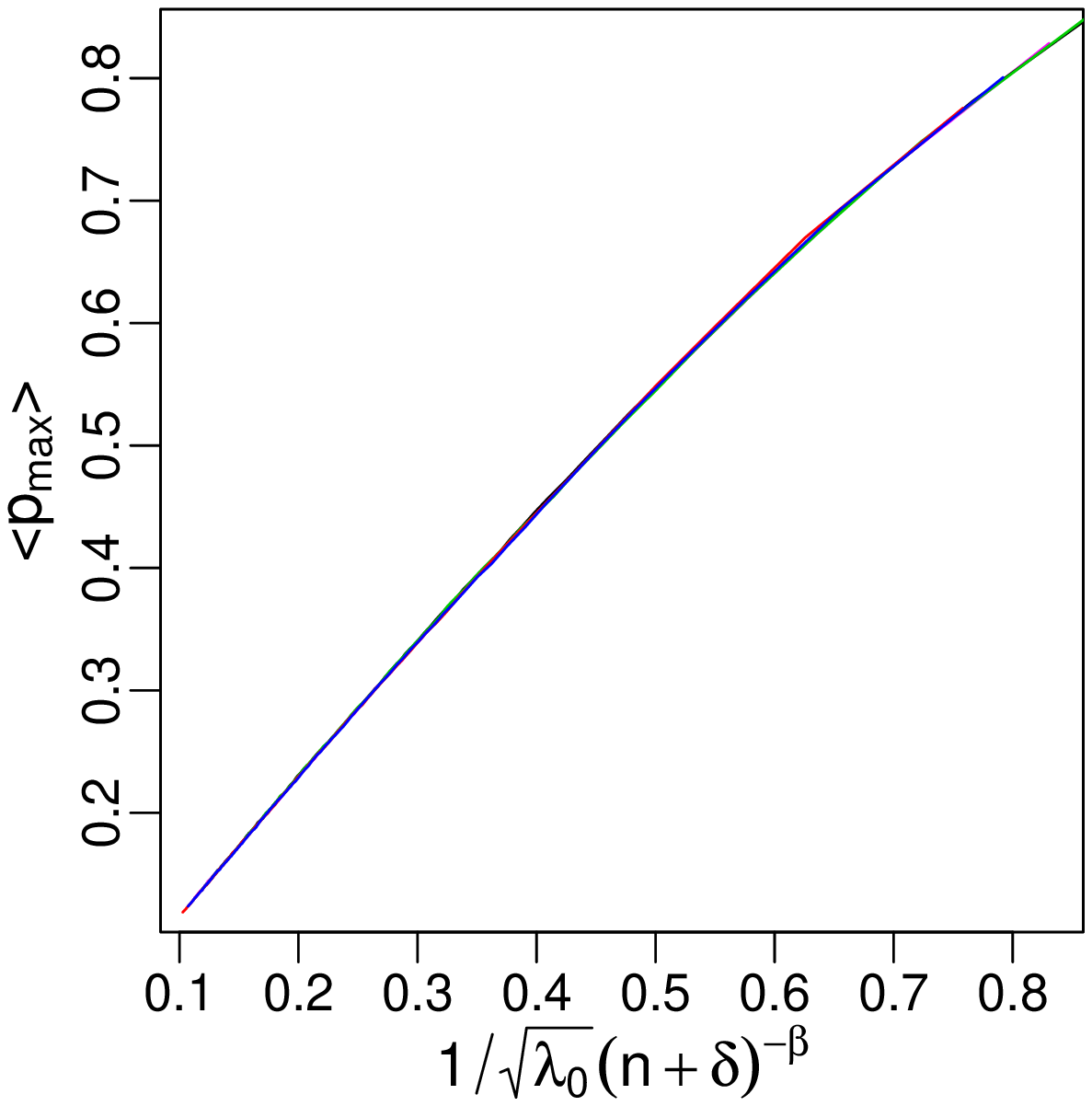}
\caption{The upper panel shows data of $\langle p_{max}\rangle$ versus
  $n$ for $\lambda_0=.8,.9,1.0,1.1,1.2$. The lower panel shows the same
  data plotted versus the scaling variable defined in Eq. (\ref{scvar}). The estimated values of
  $\beta$ and $\delta$ that lead to the best data collapse, using least squares,
  are $\beta_{max}=.389$, $\delta_{max}=.505$.  We therefore have from $D\approx 1.713$ that
$\alpha_{min}\approx 0.666$.}
\label{collapse}
\end{figure}

The direct measurement, while correct, cannot guarantee that very slow 
convergence of the power laws as a function of $n$ may somehow hide an asymptotic identity of
$\alpha_{min}$ and $\alpha_{tip}$. To remove this worry we adopt now
the scaling function technique of \cite{00DLP} to achieve an accurate determination of $\alpha_{min}$. 
In this approach one acknowledges that Eq. (\ref{pmax}) may be realized only asymptotically
for high values of $n$. For low and medium values of $n$, $\langle p_{max}\rangle$,
which is a function of the discrete $n$ and of $\lambda_0$,  
is in fact a scaling function of a single scaling variable, 
\begin{eqnarray}
\langle p_{max}\rangle= f_{\delta,\beta}(x)\ , \nonumber\\
x = \frac 1 {\sqrt\lambda_0}(n+\delta)^{-\beta} \ , \label{scvar}
\end{eqnarray}
where we have denoted $\beta=\alpha_{min}/D$.
The difference with Eq. (\ref{pmax}) is that $f_{\delta,\beta}$ is in general
not a linear function of its argument, except at exceedingly small values of $x$,
when $n$ is very large. In Fig. \ref{collapse} we demonstrate the existence of the
scaling function and the excellent data collapse achieved using it.
In the upper panel we plot $f_{\delta,\beta}(\lambda_0,n)$ for five values of
$\lambda_0$ and $1\leq n \leq 250$. In the lower panel the same data
are collapsed using the single scaling variable. We draw the reader's attention
to the following two observations: (i) the data collapse is available immediately, even for the 
smallest values of $n$ \cite{00DLP}, and (ii) the scaling function is not linear throughout
the range explored here. Thus the scaling law (\ref{pmax}) is not obeyed yet for values of $n$
of the order of a few hundreds. The set of parameters $\beta$ and
$\delta$ which give the best data collapse in the lower panel
are $\beta=.389$ and $\delta=.505$. These
parameters are used in the lower panel and give the
estimate $\alpha_{min}=0.666$ when assuming that the fractal dimension is $D=1.713$.

An even more accurate determination of $\alpha_{min}$ is achieved next.
Taking the data collapse as an evidence for the existence of a scaling
function, we conclude that for any two pairs  of numbers $(n,\lambda_0)$ and
$(\bar n,\bar\lambda_0)$ that satisfy the equation
\begin{equation}
  \label{cond}
  \frac 1 {\sqrt{\lambda_{0}}}(n+\delta)^{-\alpha_{min}/D}=  \frac 1
  {\sqrt{\tilde\lambda_{0}}}(\tilde n+\delta)^{-\alpha_{min}/D} \ ,
\end{equation} 
it follows that
\begin{equation}
  \label{feqf}
  f_{\delta,\beta}(\lambda_0,n)=f_{\delta,\beta}(\tilde\lambda_0,\tilde n) \ .
\end{equation}
These equations offer a calculational procedure. We find $\langle p_{max}\rangle$ for 
a given $n$ and $\lambda_0$, and then for another value $\bar n$ seek the value
$\bar\lambda_0$ for which $\langle p_{max}\rangle$ is the same. From Eq. (\ref{cond}) 
we then deduce that 
\begin{equation}
  \label{eq:12}
  \alpha_{min}=\frac 1 2 D \frac{\log\lambda_{0}
  -\log\tilde\lambda_{0} }{\log(n+\delta)-\log(\tilde n+\delta)}
\end{equation}
\begin{figure}
\centering
\includegraphics[width=.35\textwidth]{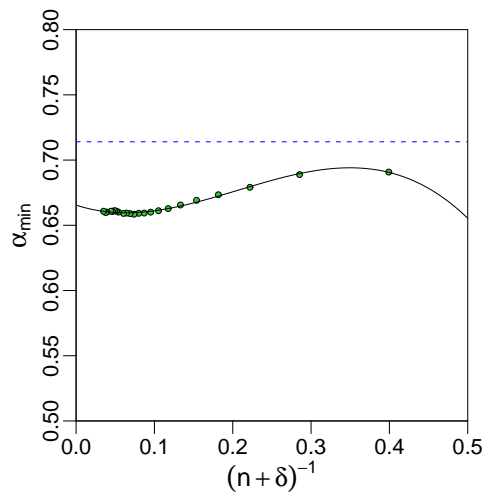}
\caption{Data of $\alpha_{min}$ estimated using Eq. (\ref{eq:12}) with
  $\tilde n=n+1$. The data is
  fitted with a cubic polynomial. The polynomial intersects the y-axis at the
  value $\alpha_{min}=0.665$. Using upper and lower values of $\delta$, $\delta=0$ and
  $\delta=1$ we estimate the following bounds on the value of
  $\alpha_{min}$, $0.662<\alpha_{min}<0.669$.}\label{extrapol}
\end{figure} 
\begin{figure}
\centering
\includegraphics[width=.35\textwidth]{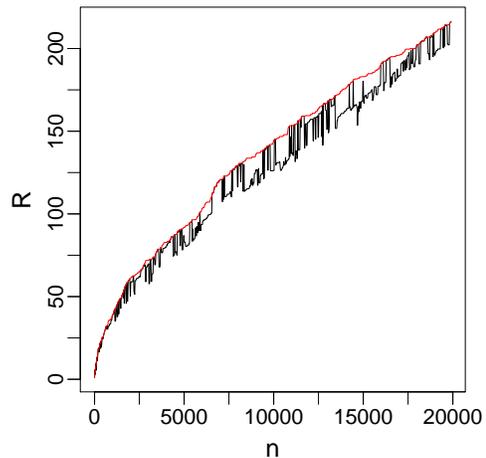}
\caption{The value of $R_n$ as a function of $n$ (upper curve) and the position
of the bump belonging to $\alpha_{min}$ (lower curve) in a typical DLA}
\label{rmax}
\end{figure}
In Fig. \ref{extrapol} we present the results of such a calculation with $\bar n=n+1$, and $1\le n
\le 250$. Since $\delta$ is not known a priori, we used the value $\delta=0.505$ which was
extracted from the data collapse in Fig. \ref{collapse}. We checked the sensitivity
to $\delta$ by bracketing the results with $\delta=0$ and $\delta=1$ respectively.
The data in Fig. \ref{extrapol} correspond to $\delta=0.505$. Fitting the data with a
cubic polynomial and extrapolating to $x\to 0$ we get the value
$\alpha_{min}\approx 0.665$. On the other hand if we repeat the procedure
using the  values of
$0\leq \delta\leq 1$ we are able to bracket the estimate in the interval
\begin{equation}
  \label{eq:15}
0.662<\alpha_{min}<0.669  \ .
\end{equation}
We thus conclude the analysis with the estimate $\alpha_{min} = 0.665\pm 0.004$.

Finally, we explain why the Turkevich-Scher conjecture (\ref{TS}) fails. The reason is that
the points corresponding to $p_{max}$ and $p_{tip}$ are not at all the same in
typical DLA. In Fig. \ref{rmax} we present the calculated value of $R_n$, computed
from the position of largest $|z|$ on the cluster, compared with the position
corresponding to the maximal harmonic measure. We see that the position of maximal
probability fluctuates wildly, and the fluctuations do not appear to go down
with the increase in the cluster size. The loss of the conjecture $(\ref{TS})$
means that there is no clear connection between the spectrum of singularities $f(\alpha)$
and the fractal dimension of DLA. As said above, the relation (\ref{tip}) is
a tautology once the existence of the scaling law (\ref{ptip}) has
been established \cite{02BS2}. Since the value of $\alpha_{tip}$ has nothing to
do with the edge of the $\alpha$ spectrum, it appears as hard to determine it
from first principles as to determine the dimension $D$ itself.

\end{document}